\documentclass[aps,prl,reprint,superscriptaddress,floatfix,longbibliography]{revtex4-2}

\usepackage{amsmath,amssymb,bm}
\usepackage{graphicx}
\usepackage{microtype}

\newcommand{\kb}{k_{\mathrm B}}
\newcommand{\dd}{\,\mathrm d}

\begin{document}

\title{Universal Entropic Occupation Statistics in Disordered Bosonic Resonators}

\author{Orest Bucicovschi}
\author{David A. Meyer}
\affiliation{Department of Mathematics, University of California San Diego, La Jolla, California 92093-0112, USA}
\email{Contact author:  dmeyer@ucsd.edu}

\date{8 August 2026}

\begin{abstract}
Programmable microcavities support grand-canonical photon gases.  We show that bosonic state counting creates an entropic staircase of 
most-probable total occupations.  For detuning density continuous and 
nonzero at the chemical-potential threshold, the active-cell fraction 
is asymptotically linear at low temperature and the conditional law 
has a universal limit; for uniform disorder the law is exact over a 
finite temperature interval.  For two modes it is the Gauss--Kuzmin 
distribution, linking photonic thermodynamics and metric number 
theory.  We outline finite-array and dye-microcavity tests.
\end{abstract}

\maketitle

Dye-filled optical microcavities make light behave, to a useful 
approximation, as a thermalized Bose gas with a controllable chemical 
potential.  Repeated absorption and emission by dye molecules can 
establish a Bose--Einstein distribution at the dye 
temperature~\cite{Klaers2010Therm,Klaers2010BEC}.  The molecular 
excitations also act as a particle reservoir:  experiments with large 
molecular reservoirs have observed grand-canonical photon-number 
fluctuations and the corresponding fluctuation--dissipation 
relation~\cite{Schmitt2014,Ozturk2023}.  Meanwhile, microstructured 
mirrors have produced programmable wells, coupled sites, and photonic 
lattices~\cite{Dung2017,Kurtscheid2020,Redmann2024}, culminating in 
the thermalization of photons in a controllable two-state system of
light~\cite{Kurtscheid2025}.  These developments make it natural to 
ask what occupation statistics arise when effectively independent 
few-mode cells sample a distribution of chemical-potential offsets, 
or detunings, from a common reservoir.

We show that the competition between occupation energy and 
bosonic-state-counting entropy produces a staircase of most-probable 
(dominant) total occupations.  The fraction of cells whose dominant 
total-occupation number is nonzero grows asymptotically 
linearly at low temperature.  Conditioned on this thermally active 
subset, the occupation histogram approaches a universal law; for 
uniform disorder the limit is reached over a finite temperature 
interval.  The two-mode law is the Gauss--Kuzmin distribution of 
metric number theory~\cite{Khinchin1997}.

\begin{figure}[ht]
\centering
\includegraphics[width=\columnwidth]{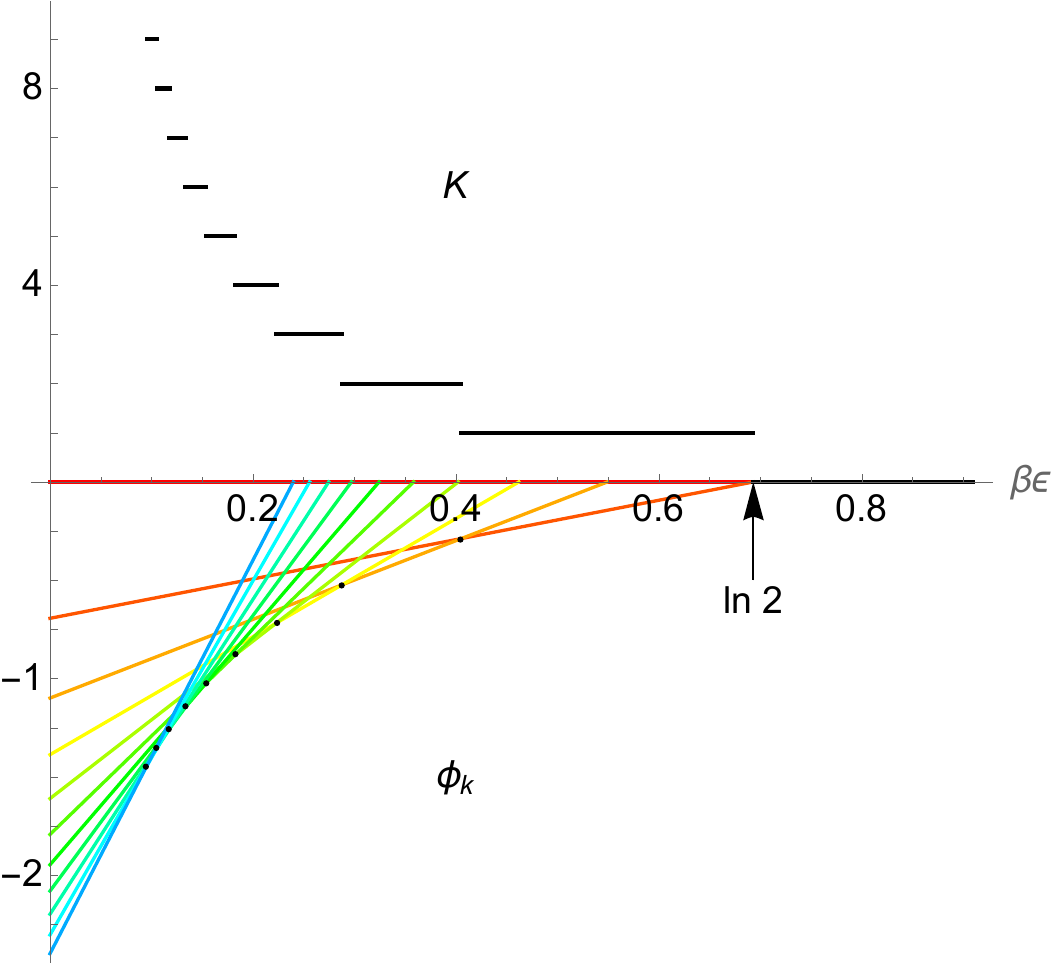}
\caption{Dominant total occupation from the occupation-sector grand 
potentials of a two-mode cell.  Below the horizontal axis, the 
straight lines are $\phi_k = kx-\ln(k+1)$, with $x = \beta\epsilon$ 
and $k=0,\ldots,10$; black dots mark successive crossings of the 
lower envelope.  Above the axis, the same abscissa carries the 
staircase $K(x)$.  The ordinate is split:  it denotes $K$ above zero 
and $\phi_k = \beta\Phi_k$ below zero.  The $K = 0$ plateau lies on 
the horizontal axis for $x > \ln 2$.}
\label{fig:mechanism}
\end{figure}

\textit{Grand-canonical two-mode cells---}%
Consider $N$ independent cells.  Cell $i$ contains two degenerate, 
noninteracting bosonic modes of energy $E_i$, with negligible hopping 
or interaction between cells.  The cells exchange particles and energy 
with a reservoir large enough that its temperature $T$ and chemical
potential $\mu$ remain fixed.  Define the detunings 
$\epsilon_i = E_i - \mu > 0$, and let $x_i = \beta\epsilon_i$, where 
$\beta = (\kb T)^{-1}$.  The detunings are quenched, \textit{i.e.}, 
fixed by fabrication or external control throughout equilibration
and measurement, but varying from cell to cell.  For a representative 
cell we suppress the index and write $\epsilon = \epsilon_i$ and 
$x = x_i$.  Then for a single cell the joint occupation probability is
$P_{\epsilon}(n_1,n_2) = e^{-x n_{\rm tot}}/\Xi_\epsilon$, where 
$\Xi_\epsilon = (1-e^{-x})^{-2}$ is the grand partition function and 
$n_{\rm tot} = n_1 + n_2$ is the fluctuating total photon occupation 
of the cell.  For $k\in\{0,1,2,\ldots\}$, summing the probabilities of 
the $k+1$ nonnegative occupation pairs $(n_1,n_2)$ with 
$n_{\rm tot} = k$ gives
\begin{equation}
  P_{\epsilon}(k)
  = 
  \Pr(n_{\rm tot} = k\mid\epsilon)
  =
  (1 - e^{-x})^2 w_k(x),
 \label{eq:Pk}
\end{equation}
where $w_k(x) = (k+1)e^{-kx}$ is the unnormalized grand-canonical 
sector weight.  The corresponding occupation-sector grand potential is
$\Phi_k(\epsilon,T) = -\kb T\ln w_k(\beta\epsilon)$, which has 
dimensionless form:
\begin{equation}\label{eq:restricted-potential}
  \phi_k(x) = \beta\Phi_k(\epsilon,T) = kx - \ln(k+1).
\end{equation}
Except at the isolated crossing points, the most-probable occupation 
number is $K(x) = \operatorname*{arg\,min}_{k\geq0}\phi_k(x)$.  The 
linear energy penalty and logarithmic state-counting entropy generate 
the crossings shown in Fig.~\ref{fig:mechanism}.

Now, for $k\ge 1$,
\begin{equation}
  \frac{P_{\epsilon}(k)}{P_{\epsilon}(k-1)} = \frac{k+1}{k}e^{-x},
\end{equation}
so setting $b_k = \ln\bigl((k+1)/k\bigr)$, we have, away from isolated
ties,
\begin{equation}\label{eq:plateau}
  K(x) = k
  \quad\Longleftrightarrow\quad
  b_{k+1} < x < b_k.
\end{equation}
The first crossing is at $b_1 = \ln 2$:  a cell is \textsl{active}, 
meaning $K\ge 1$, precisely when
\begin{equation}\label{eq:active}
  0 < \epsilon < \kb T\ln2.
\end{equation}
At this boundary the energy cost of one boson is balanced by the 
entropy $\kb\ln2$ of choosing between the two modes.

\textit{Universal active-cell law---}%
Let the quenched detunings be distributed according to a probability 
density $\rho(\epsilon)$ for $\epsilon > 0$, and put $t = \kb T$.  By 
Eq.~\eqref{eq:active}, the fraction of cells that are active, 
\textit{i.e.}, have $K\ge 1$, is
\begin{equation}\label{eq:activefraction}
  A(T)
  = \Pr(K\ge 1)
  = \int_0^{t\ln2}\!\rho(\epsilon)\dd\epsilon.
\end{equation}
For $k\ge 1$, the probability that an active cell has dominant 
total-occupation sector $k$ is
\begin{equation}\label{eq:generalpk}
  p_k(T)
  = \Pr(K = k\mid K\ge 1)
  = \frac{1}{A(T)}
    \int_{t b_{k+1}}^{t b_k}\rho(\epsilon)\dd\epsilon.
\end{equation}
If $\rho$ is continuous at the chemical-potential threshold
$\epsilon = 0$ and $\rho(0) > 0$, then, as $T\to 0$,
\begin{align}
  A(T)
  &= \rho(0)t\ln2\,[1+o(1)],                      \label{eq:Alinear}\\
  p_k(T)
  &\longrightarrow
  p_k^{(2)}
  =
  \frac{b_k-b_{k+1}}{\ln2}
  =
  \frac{1}{\ln2}
  \ln\frac{(k+1)^2}{k(k+2)},                          \label{eq:GK}
\end{align}
for $k\geq1$.  The limit $p_k^{(2)}$ is exactly the
Gauss--Kuzmin distribution~\cite{Khinchin1997}.  The
identification is more than an equality of probability functions:  
Under the change of variable $\xi = e^{\beta\epsilon}-1$, the 
conditional active-cell measure tends to the invariant Gauss measure
for $0 < \xi < 1$,
$$
  \dd\mu_{\rm G}(\xi)
  =
  \frac{\dd\xi}{\ln2\,(1+\xi)},
$$
while Eq.~\eqref{eq:plateau} becomes $K = \lfloor1/\xi\rfloor$ away 
from the measure-zero boundaries.  Hence
$$
  p_k^{(2)}
  =
  \mu_{\rm G}
  \left(
    \frac{1}{k+1},\frac{1}{k}
  \right],
$$
the Gauss--Kuzmin digit probability.  For uniform disorder this 
correspondence is exact throughout the finite temperature interval 
considered below.  The Gauss map is the first-return map induced by 
the Farey map~\cite{Isola2002}, whose transfer-operator thermodynamics 
is closely connected with Farey-fraction spin 
chains~\cite{KlebanOzluk1999,PrellbergFialaKleban2006}.  In those 
models Farey number-theoretic structure is built into the energy
function, whereas here the Gauss measure emerges from bosonic state 
counting and detuning disorder.  The resulting Gauss--Kuzmin law has 
survival probabilities
\begin{equation}\label{eq:tail}
  S_k^{(2)}
  =
  \sum_{j=k}^{\infty}p_j^{(2)}
  =
  \frac{b_k}{\ln2}
  =
  \log_2\left(1+\frac1k\right),
  \qquad k\geq1,
\end{equation}
which decays as $1/k$.

\begin{figure*}[thb]
\centering
\includegraphics[width=0.485\textwidth]{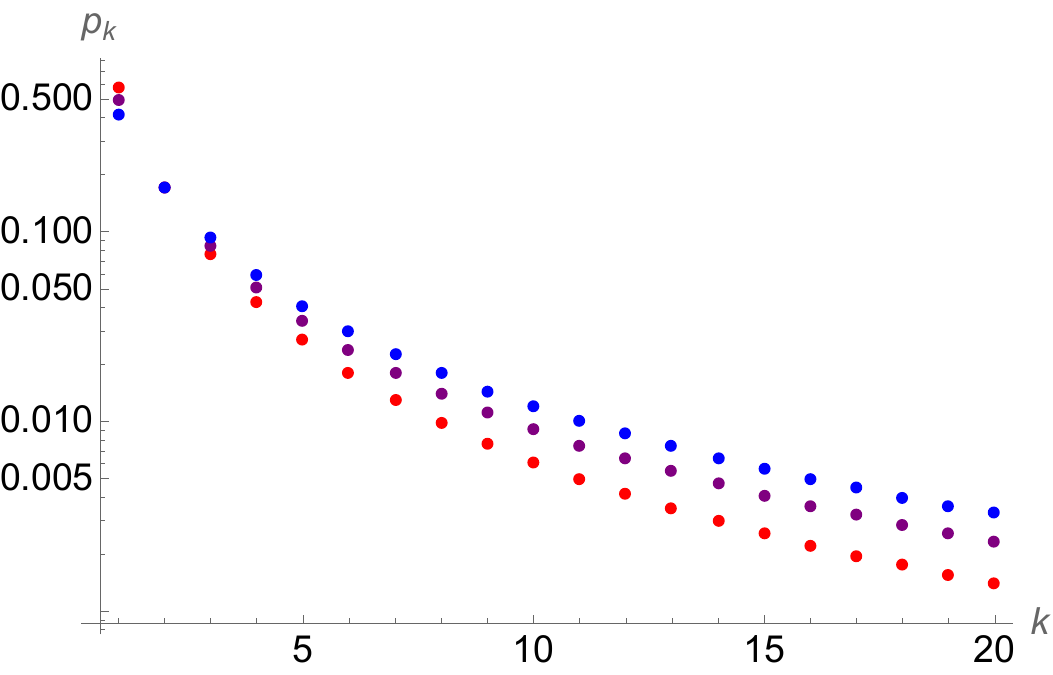}\hfill
\includegraphics[width=0.485\textwidth]{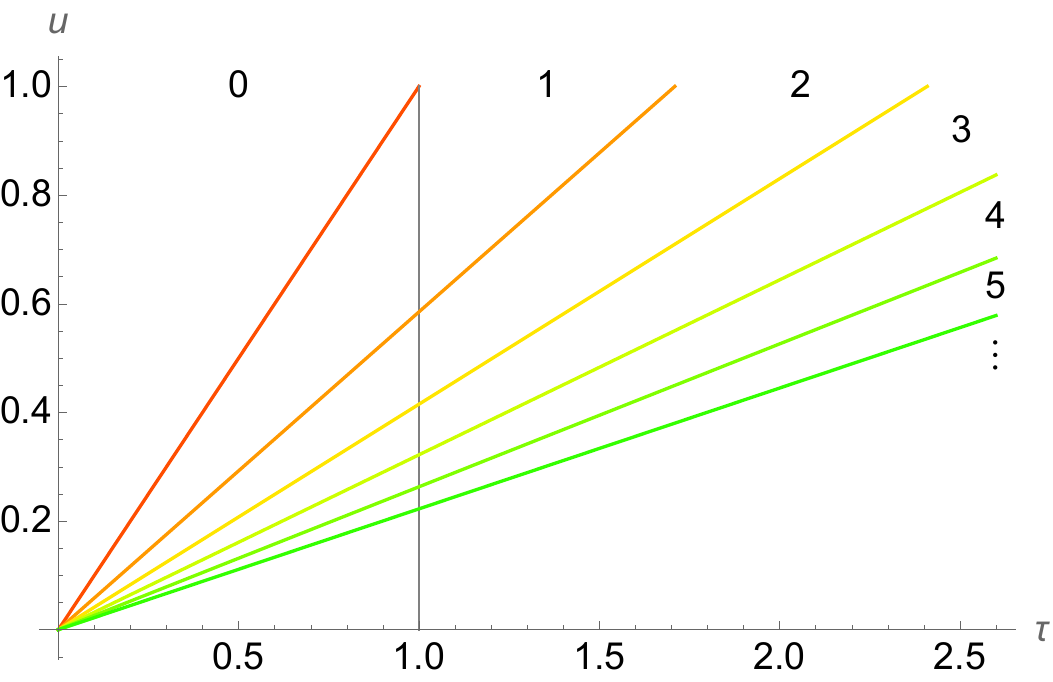}
\caption{Temperature and disorder dependence of the dominant 
occupation number.
(Left) Conditional active-cell distributions on a logarithmic scale.  
The reference distribution is $p_k^{(2)}$, obtained exactly for 
uniform disorder at every $0 < \tau \le1$ and shown in blue.  The red 
and purple points show the probability distributions for a 
representative nonuniform detuning density 
$\rho(\epsilon) = (1 + 4\epsilon/\Delta)/(3\Delta)$ at temperatures 
$\tau = 1$ and $\tau = 1/4$, respectively, approaching $p_k^{(2)}$ as 
$T \to 0$.
(Right) Dominant-occupation-number sector map for a uniform detuning 
band, in $u=\epsilon/\Delta$ and $\tau=T/T_*$.  The colored lines 
$u_k(\tau) = \tau b_k/\ln 2$ separate sectors $K = k-1$ and $K = k$,
labeled by the black numerals $0,1,2,3,4,5,\ldots$.  The activation 
boundary is $u_1 = \tau$, and $\tau = 1$ is the entropy--bandwidth 
matching point.  Because $u$ is uniformly distributed, the vertical 
width of each sector equals its population fraction at that value of
$\tau$.}
\label{fig:predictions}
\end{figure*}

If, in addition, the threshold density has the expansion
$$
\rho(\epsilon)
 =
\rho(0) + \rho'(0)\epsilon + O(\epsilon^2)
$$
as $\epsilon\searrow 0$, then the leading finite-temperature 
deformation is
\begin{equation}\label{eq:correction}
  p_k(T) 
  = 
  p_k^{(2)}\bigg[1 + \frac{\rho'(0)}{2\rho(0)}t
                     \big(b_k+b_{k+1}-\ln2\big)
                   + O(t^2)
           \bigg].
\end{equation}
Thus the first departure from the universal law is controlled by the 
local logarithmic slope, $\rho'(0)/\rho(0)$, of the detuning density 
at threshold.  The left panel of Fig.~\ref{fig:predictions} 
illustrates this deformation for a linear detuning density and its 
approach to $p_k^{(2)}$ as $T\to 0$.  Appendix A proves that the full 
discrete distribution converges in total variation.

The two-mode law is the first nontrivial member of a mode-degeneracy 
family.  For $q\geq2$ degenerate modes, there are $\binom{k+q-1}{q-1}$ 
Fock states with total occupation $k$, so Eq.~\eqref{eq:Pk} becomes
$$
  P_{\epsilon}^{(q)}(k)
  =
  \binom{k+q-1}{q-1}
  (1-e^{-x})^q e^{-kx},
$$
for $k = 0,1,\ldots$.  The adjacent-sector ratio is
$$
  \frac{P_{\epsilon}^{(q)}(k)}
       {P_{\epsilon}^{(q)}(k-1)}
  =
  \frac{k+q-1}{k}e^{-x},
$$
so the boundary between sectors $k-1$ and $k$ is
$$
  c_k^{(q)}
  =
  \ln\frac{k+q-1}{k}.
$$
In particular, $c_1^{(q)} = \ln q$, and the active window has width 
$\kb T\ln q$.  The same threshold-rescaling argument therefore gives, 
for $k\ge 1$,
\begin{equation}\label{eq:qmode}
  p_k^{(q)}
  =
  \frac{c_k^{(q)}-c_{k+1}^{(q)}}{\ln q}
  =
  \frac{1}{\ln q}
  \ln\frac{(k+q-1)(k+1)}{k(k+q)}.  
\end{equation}
For integer $q\ge 2$, these are also the invariant-measure
probabilities of the reciprocal intervals $(1/(k+1),1/k]$ for the 
$(q-1)$-simple continued-fraction map~\cite{FernandezSanchezEtAl2023}; 
at $q=2$ this reduces to the Gauss measure and the ordinary 
Gauss--Kuzmin law.

\textit{Uniform disorder and an exact finite temperature law---}%
For a uniform band $0 < \epsilon < \Delta$, define
\begin{equation}\label{eq:Tstar}
  T_* = \frac{\Delta}{\kb\ln2}
  \text{\ \ and\ \ }
  \tau = \frac{T}{T_*}.
\end{equation}
For every $0 < \tau \le 1$,
\begin{equation}\label{eq:zeroinflated}
  \Pr(K = k)
  =
  \begin{cases}
    1-\tau         &\text{if\ } k = 0; \\
    \tau p_k^{(2)} &\text{if\ } k\ge1.
  \end{cases}
\end{equation}
The active fraction grows linearly with $T$, while the conditional 
occupation distribution is independent of temperature and equals 
Eq.~\eqref{eq:GK} for every $0 < T\le T_*$.  At $T = T_*$ the
inactive fraction vanishes, and the unconditional histogram is 
Gauss--Kuzmin.  The condition $\beta\Delta = \ln2$ is thus an 
entropy--bandwidth matching point, not a singular thermodynamic 
transition.  At arbitrary temperature,
\begin{equation}\label{eq:arbitrary-temperature}
  \Pr(K\ge k)
  =
  \min\left\{1,\frac{b_k}{\beta\Delta}\right\}, \text{\ for\ }k\ge 1. 
\end{equation}
The corresponding map of the most-probable total-occupation sector in 
the $(\tau,u)$ plane, with $u = \epsilon/\Delta$, is shown in the 
right panel of Fig.~\ref{fig:predictions}.  Above $T_*$, the 
low-occupation sectors disappear successively as the upper edge of the 
detuning band crosses the staircase boundaries.

\textit{Finite cells and experimental protocol---}%
No thermodynamic limit is required for the law itself.  For $N$ cells 
with independently sampled uniform detunings and $T\le T_*$, the 
active-cell count satisfies 
$N_{\rm a}\sim\operatorname{Binomial}(N,\tau)$.  Conditional on 
$N_{\rm a}$, the positive-sector counts are multinomial with 
probabilities $p_k^{(2)}$, so sampling errors in any fixed sector or 
survival probability scale as $N_{\rm a}^{-1/2}$.  For $N\tau\gg1$, 
the typical scale is therefore $(N\tau)^{-1/2}$.  At $T = T_*$, 
setting the detunings to the midpoints of $N$ equal subintervals 
reduces the survival function discrepancy to at most $1/(2N)$ (see 
Appendix B).  The ideal $1/k$ survival tail makes the largest dominant 
total occupation among $N_{\rm a}$ active cells typically 
$O(N_{\rm a})$.  Physical nonidealities regularize the far tail; for 
small mode splitting and a large reservoir, the low-$k$ sector 
probabilities remain perturbatively close to the ideal law.

A direct realization would use an array of spatially separated or 
negligibly coupled dye-microcavity cells, each supporting two nearly 
degenerate spatial modes of one selected polarization; the orthogonal 
polarization should be energetically or dissipatively suppressed 
because polarization thermalization need not be 
complete~\cite{Moodie2017}.  The mean cell energy would be set so 
that $E_i - \mu$ samples the desired density.  Existing experiments 
separately demonstrate grand-canonical number statistics and the 
corresponding fluctuation--dissipation 
behavior~\cite{Schmitt2014,Ozturk2023}, programmable and coupled 
trapping potentials~\cite{Dung2017,Kurtscheid2020,Redmann2024},
and a thermalized two-spatial-mode system with splitting far below 
$\kb T$~\cite{Kurtscheid2025}.  At room temperature, 
$T = 300\,\mathrm{K}$, the active two-mode bandwidth is 
$\kb T\ln2\simeq 17.9\,\mathrm{meV}$.  A single programmable cell 
swept through a detuning grid reproduces the same ensemble histogram, 
provided it equilibrates at each setting.

The dominant occupation-number sector is also experimentally simpler 
than its definition might suggest.  In the ideal two-mode 
grand-canonical model,
\begin{equation}\label{eq:meanK}
  \overline n_{\mathrm{tot}}
  = 
  \frac{2}{e^{\beta\epsilon}-1} \text{\ and\ }
  K = \left\lfloor\frac{\overline n_{\mathrm{tot}}}{2}\right\rfloor, 
\end{equation}
away from the measure-zero plateau boundaries.  Thus calibrated mean 
emitted intensity is sufficient to assign $K$; full counting 
statistics provide a stronger test but are not required.

Finally, the equilibrium variable $x = \beta\epsilon$ has a direct 
open-system calibration.  For a fixed molecular reservoir, birth and 
death rates $\gamma_+(n+1)$ and $\gamma_-n$ give a stationary 
geometric mode distribution with ratio $r = \gamma_+/\gamma_-$.  Two 
equivalent modes therefore reproduce Eq.~\eqref{eq:Pk} with
\begin{equation}\label{eq:effective_x}
  x = -\ln r.
\end{equation}
In equilibrium, detailed balance gives $x = \beta\epsilon$; 
absorption, emission, and cavity loss can instead be folded into the 
measured $r$ and compensated by programming a grid uniform in 
$x$~\cite{Kirton2013,Kirton2015}.  This fixed-reservoir model provides 
a direct open-system realization and calibration of the ideal variable 
$x$.  Finite-reservoir effects are parametrically small in the
large-reservoir, low-occupation regime and primarily regularize the 
far tail (see Appendix B).

We have identified an entropic occupation staircase and a universal 
conditional law in disordered grand-canonical bosonic resonators.  Its 
linked signatures are asymptotically linear low-temperature growth of 
the active-cell fraction, an exact finite-temperature conditional law 
for uniform disorder, a deformation of the law controlled by the local
logarithmic slope of the detuning density at threshold, and 
finite-size scaling in the active population.  The Gauss--Kuzmin 
identification is the two-mode expression of this broader statistical 
mechanical mechanism.

\textit{Data availability---}%
No data were generated or analyzed for this theoretical study.

\bibliography{entropicoccupation}

\setcounter{equation}{0}
\renewcommand{\theequation}{A\arabic{equation}}

\vskip\topsep
\noindent\textit{Appendix A:  Threshold density expansion and finite
temperature correction---}%
Put $t = \kb T$ and $L = \ln 2$.  Conditioned on activity, the scaled 
detuning $0 < y = \epsilon/(tL) < 1$ has density
\begin{equation}\label{eq:scaled_density}
  f_t(y) = \frac{tL\rho(tLy)}{\int_0^{tL}\rho(\epsilon)\dd\epsilon}.
\end{equation}
For $\rho(\epsilon) = \rho_0 + \rho_1\epsilon + O(\epsilon^2)$, where 
$\rho_0 = \rho(0)$ and $\rho_1 = \rho'(0)$,
\begin{equation}\label{eq:scaled_expansion}
  f_t(y) 
  =
  1 + \frac{\rho_1}{\rho_0}tL\left(y - \frac12\right) + O(t^2).
\end{equation}
Occupation sector $k\ge 1$ is the interval $I_k = (b_{k+1}/L,b_k/L)$.  
Integrating Eq.~\eqref{eq:scaled_expansion} over $I_k$ gives 
Eq.~\eqref{eq:correction}.  In fact, the total-variation distance from
$p_k^{(2)}$ can be bounded:
\begin{equation}\label{eq:TV}
  \frac{1}{2}\sum_{k\geq1}|p_k(T)-p_k^{(2)}|
  \le
  \frac{1}{2}\int_0^1|f_t(y)-1|\dd y.
\end{equation}
Then continuity of $\rho$ at zero with $\rho(0) > 0$ implies 
$f_t\to 1$ uniformly on $[0,1]$, so the total-variation distance tends 
to zero.

For $q$ modes the restricted multiplicity is 
$g_k^{(q)} = \binom{k+q-1}{q-1}$, and the boundary between sectors 
$k-1$ and $k$ is
\begin{equation}
  c_k^{(q)} = \ln\frac{k+q-1}{k}.
\end{equation}
The active window has width $t\ln q$, and the same rescaling proof 
gives Eq.~\eqref{eq:qmode} and the survival law
$$
  S_k^{(q)}
  =
  \frac{c_k^{(q)}}{\ln q}
  =
  \frac{1}{\ln q}
  \ln\left(1+\frac{q-1}{k}\right), \text{\ for\ }k\ge 1.
$$

\setcounter{equation}{0}
\renewcommand{\theequation}{B\arabic{equation}}

\vskip\topsep
\noindent\textit{Appendix B: Experimental nonidealities and 
finite-cell statistics---}%
For one mode coupled to a reservoir whose molecular populations are
effectively fixed, consider the birth--death process
\begin{equation}
  n\mapsto
  \begin{cases}
    n+1 &\text{at rate } \gamma_+(n+1),\\
    n-1 &\text{at rate } \gamma_-n.
  \end{cases}
\end{equation}
The flux balances when $\Pr(n+1)/\Pr(n) = r = \gamma_+/\gamma_-$;  
hence $\Pr(n) = (1-r)r^n$ for $r < 1$.  For two equivalent, 
noninteracting modes with independent reservoir-induced transitions, 
the stationary joint distribution factorizes as
\begin{equation}
  \Pr(n_1,n_2) = (1-r)^2r^{n_1 + n_2}.
\end{equation}
Writing $n_{\mathrm{tot}} = n_1 + n_2$ for the instantaneous total
occupation and summing over the $k+1$ pairs with $n_1 + n_2 = k$ gives
\begin{equation}
  \Pr(n_{\mathrm{tot}} = k) = (k+1)(1-r)^2r^k.
\end{equation}
Comparison with Eq.~\eqref{eq:Pk} identifies $r = e^{-x}$, or
equivalently $x = -\ln r$, as in Eq.~\eqref{eq:effective_x}.  Cavity
loss may be included additively in $\gamma_-$; the result remains 
exact provided the molecular populations are effectively fixed.

As a simple estimate of finite-reservoir corrections, suppose that the
reservoir contains $M$ two-level dye molecules and that a fixed total
number $X$, with $0 < X < M$, of excitations is shared between the
molecules and the two cavity modes.  If the modes contain $k$ photons
in total, with $0\leq k\leq X$, then $X-k$ molecules are 
electronically excited and $M-X+k$ are in their ground state.  The 
molecular multiplicity of this sector is therefore $\binom{M}{X-k}$.

Let $Y = M-X$, the number of ground-state molecules in the zero photon
sector.  Relative to that sector, the molecular multiplicity satisfies
\begin{equation}
  \frac{\binom{M}{X-k}}{\binom{M}{X}}
  =
  \prod_{j=0}^{k-1}\frac{X-j}{Y+j+1}.
\end{equation}
When $k/X\ll 1$ and $k/(Y+1)\ll 1$, expanding the logarithm in the
small ratios $j/X$ and $j/(Y+1)$ gives
\begin{align}\label{eq:finite-reservoir}
  \ln\frac{\binom{M}{X-k}}{\binom{M}{X}}
  ={}&
  k\ln\frac{X}{Y+1}
  -\frac{k(k-1)}{2}\left(\frac{1}{X}+\frac{1}{Y+1}\right)    \notag\\     
  &+
  O\left(
    \frac{k^3}{X^2} + \frac{k^3}{(Y+1)^2}
  \right).
\end{align}
The first term, which is linear in $k$, renormalizes the effective
fugacity, whereas the $k(k-1)$ term is the leading deformation from
grand-canonical statistics.  Setting 
$M_{\rm eff}^{-1} = X^{-1} + (Y+1)^{-1}$, 
Eq.~\eqref{eq:finite-reservoir} shows that finite-reservoir 
corrections are negligible for $k^2\ll M_{\rm eff}$.  
Experiments with large dye reservoirs have directly observed 
grand-canonical number statistics~\cite{Schmitt2014}, while related
work has verified the fluctuation--dissipation relation in the same 
platform~\cite{Ozturk2023}.  Finite-reservoir capacity therefore 
provides one physical regularization of the far tail but is not needed 
to describe the first several sectors.

For two modes split by $\delta\ge 0$, with detunings $\epsilon$ and
$\epsilon + \delta$, put $z = \beta\delta$.  In the total-occupation 
sector $n_{\mathrm{tot}} = k$, let $m$ denote the occupation of the 
higher energy mode.  The splitting weighted multiplicity is
\begin{equation}
  h_k(z) = \sum_{m=0}^{k}e^{-mz} = \frac{1-e^{-(k+1)z}}{1-e^{-z}},
\end{equation}
where the final expression is understood by continuity to mean 
$h_k(0) = k+1$.  The $k-1\leftrightarrow k$ boundary is
\begin{equation}\label{eq:split_boundary}  
  x_k(z)
  =
  \ln\frac{h_k(z)}{h_{k-1}(z)}
  =
  \ln\frac{1-e^{-(k+1)z}}{1-e^{-kz}}.
\end{equation}
For fixed $k$ and $z\ll 1$,
\begin{equation}
  x_k(z) = b_k - \frac{z}{2} + \frac{2k+1}{24}z^2 + O(z^4).
\end{equation}
In particular, the active window narrows linearly,
\begin{equation}
  x_1(z) = \ln(1+e^{-z}) = \ln2 - \frac{z}{2} + O(z^2).
\end{equation}
By contrast, the width $x_k(z) - x_{k+1}(z)$ of any fixed low-$k$ 
plateau has no linear shift.  For $kz\gtrsim1$, the far-tail plateau 
widths are exponentially suppressed.

Finally, consider the uniform detuning model with $T\leq T_*$.
Conditional on having $N_{\mathrm a}$ independent active cells, let
$H_k = |\{1\leq i\leq N_{\mathrm a}\mid K_i = k\}|$, for $k\ge 1$.  
The sector counts $(H_1,H_2,\ldots)$ have the multinomial law with 
probabilities $p_k^{(2)}$.  Define the empirical survival probability 
by
\begin{equation}
  \widehat S_{N_{\mathrm a}}(k)
  =
  \frac{1}{N_{\mathrm a}}
  |\{1\le i\le N_{\mathrm a}\mid K_i\ge k\}|,
\end{equation}
for $k\ge 1$.  Then, conditional on $N_{\mathrm a}$,
\begin{equation}
  \mathsf{Var}
  \left[\widehat S_{N_{\mathrm a}}(k)\mid N_{\mathrm a}\right]
  =
  \frac{S_k(1-S_k)}{N_{\mathrm a}}
\end{equation}
where $S_k = S_k^{(2)}$ from Eq.~\eqref{eq:tail}.

At $T = T_*$, set the $N$ cell detunings to the midpoints of
$N$ equal subintervals of $(0,\Delta)$ so that 
$\epsilon_i = \Delta(i-\tfrac12)/N$, for $i = 1,\ldots,N$.  All these 
cells are active, so $N_{\mathrm a} = N$.  This deterministic midpoint 
discretization of the uniform detuning distribution obeys
\begin{equation}
  \sup_{k\ge 1}
  \left|
    \widehat S_N(k)-S_k
  \right|
  \le \frac{1}{2N},
\end{equation}
whereas independently sampled uniform detunings have typical
empirical fluctuations of order $N^{-1/2}$.

\end{document}